\documentclass[conference]{IEEEtran}
\usepackage[utf8]{inputenc}
 
\usepackage{cite}
\usepackage{amsmath,amssymb,amsfonts}
\usepackage{graphicx}
\usepackage{textcomp}
\usepackage{bmpsize}
\usepackage{xcolor}
\usepackage{lipsum}

\usepackage{algorithm}
\usepackage{algpseudocode}

\usepackage{array}

\usepackage{tabularx}
\usepackage{subcaption}
\newcommand{\comment}[1]{}
 
\usepackage{makecell}
\usepackage{booktabs}
\usepackage{multirow}
\usepackage{cite}
\usepackage{float}
\usepackage[top=0.75in, bottom=1.1in, left=0.63in, right=0.64in, columnsep=0.25in]{geometry}
\begin{document}

\IEEEoverridecommandlockouts

\title{
Coordinated Multi-Domain Deception: A Stackelberg Game Approach
\thanks{Research was sponsored by the DEVCOM Army Research Laboratory and was accomplished under Cooperative Agreement Numbers W911NF-23-2-0012. The views and conclusions contained in this document are those of the authors and should not be interpreted as representing the official policies, either expressed or implied, of the Army Research Laboratory or the U.S. Government. The U.S. Government is authorized to reproduce and distribute reprints for Government purposes, notwithstanding any copyright notation herein.}
}

%\author{\IEEEauthorblockN{Anonymous Author(s)}}
\author{
    \IEEEauthorblockN{Md Abu Sayed\IEEEauthorrefmark{1}, Asif Rahman\IEEEauthorrefmark{1},Ahmed Hemida\IEEEauthorrefmark{2} Christopher Kiekintveld\IEEEauthorrefmark{1}, Charles Kamhoua\IEEEauthorrefmark{2}}
    
    \IEEEauthorblockA{\IEEEauthorrefmark{1}University of Texas at El Paso, El Paso, Texas 79968, USA}
    
    \IEEEauthorblockA{\IEEEauthorrefmark{2}DEVCOM Army Research Laboratory, MD 20783, USA}
    \IEEEauthorblockA{msayed@miners.utep.edu, arahman3@miners.utep.edu, ahmed.h.hemida.ctr@army.mil, cdkiekintveld@utep.edu,\\ charles.a.kamhoua.civ@army.mil}
}

\maketitle
\begin{abstract}
%This paper examines the role of coordinated deception in cybersecurity by integrating synchronized deceptive strategies across cyber and physical layers to mislead attackers and strengthen defense mechanisms. By establishing a one-to-one mapping between the two layers, defenders can deploy multi-layered deception to disrupt attacker reconnaissance at early stages and enhance threat detection. To this end, we propose a Stackelberg game model to capture the strategic interaction between attackers and defenders, leveraging Common Vulnerability Scoring System (CVSS)-based exploit probabilities and reward functions to optimize deception strategies. Our framework derives realistic reward values using security domain expertise on Common Vulnerabilities and Exposures (CVEs) from the National Vulnerability Database (NVD). Finally, we compare the proposed deception technique against state-of-the-art strategies, demonstrating its practical applicability in real-world cybersecurity scenarios.

This paper explores coordinated deception strategies by synchronizing defenses across coupled cyber and physical systems to mislead attackers and strengthen defense mechanisms. We introduce a Stackelberg game framework to model the strategic interaction between defenders and attackers, where the defender leverages CVSS-based exploit probabilities and real-world vulnerability data from the National Vulnerability Database (NVD) to guide the deployment of deception. Cyber and physical replicas are used to disrupt attacker reconnaissance and enhance defensive effectiveness. We propose a CVE-based utility function to identify the most critical vulnerabilities and demonstrate that coordinated multilayer deception outperforms single-layer and baseline strategies in improving defender utility across both CVSS versions.

%This paper investigates the role of coordinated deception in cybersecurity by integrating synchronized deceptive strategies across coupled cyber and physical systems to mislead attackers and enhance defense mechanisms. We propose a Stackelberg game-theoretic framework to model the strategic interaction between attackers and defenders, where the defender acts as the leader and the attacker as the follower. The framework leverages Common Vulnerability Scoring System (CVSS) v2 and v3 scores to derive realistic reward functions based on Common Vulnerabilities and Exposures (CVEs) from the National Vulnerability Database (NVD). By employing a one-to-one correspondence between cyber and physical layer replicas, our approach disrupts early-stage attacker reconnaissance and exploits interdependencies between layers to strengthen deception effectiveness. We formulate the problem of identifying the most critical vulnerability using a CVE-based utility function and evaluate our approach against baseline strategies, including greedy and uniformly random methods. Comparative analysis demonstrates that the proposed multilayer deception strategy yields higher defender utility than deploying deception independently within a single layer. Our results are consistent across CVSS versions, demonstrating the practical effectiveness of our coordinated deception framework in real-world cyber-physical security scenarios.

\end{abstract}


\section{Introduction}

Cyber deception is a proactive defense strategy that leverages misinformation and ambiguity to counter information asymmetry and mislead attackers, diverting them from critical assets while enhancing system resilience and early threat detection. Game theory complements this approach by modeling attacker-defender interactions in uncertain environments, enabling defenders to anticipate adversarial moves and strategically deploy deception. Together, these frameworks transform cybersecurity into a dynamic, adaptive system that increases the cost and complexity of cyberattacks while optimizing defensive decision-making through structured, predictive analysis \cite{sayed2022cyber, rahman2025using}.

Early reconnaissance is a critical phase in cyberattacks, where adversaries gather intelligence on system vulnerabilities and entry points. Multilayer deception can counter this by embedding misleading indicators across both cyber and physical layers through fake network topologies, decoy services, and manipulated system responses in cyberspace, and through physical decoys, camouflage, or misleading signals in the real environment \cite{asghar2024scalable}. These integrated tactics distort the attacker’s perception of operational structures and high-value targets, forcing them to expend resources validating unreliable data. By constructing a unified and deceptive view across both domains, defenders heighten ambiguity, disrupt cross-domain inference, and significantly increase attacker uncertainty, delaying or deterring further progression.

Such coordination requires careful planning and synchronization to ensure consistency between the cyber and physical layers and to prevent contradictions that could compromise the deception. This paper investigates multi-domain coordinated deception strategies that operate concurrently. Specifically, we develop and evaluate deception policies tailored to this multi-domain environment, leveraging the interdependencies between layers to enhance overall defensive effectiveness. Our contributions are summarized as follows:

\begin{itemize}
  \item We develop a utility function grounded in CVE identifiers and informed by CVSS v2 and v3 scoring metrics to evaluate impacts and prioritize vulnerabilities.
  \item We propose a Stackelberg game-theoretic framework for coordinated multi-domain deception that leverages cyber and physical layer replicas to disrupt early attacker reconnaissance and formulate a Mixed-Integer Quadratic Programming (MIQP) to calculate the Stackelberg equilibrium.
  \item We formulate the problem of identifying the most critical vulnerability in multilayer environment involving a strategic attacker, guided by a CVE-based utility function, and propose an effective method to solve it.
  \item We compare the defender's utility achieved by our approach against greedy and uniformly random baselines, analyzing the impact of uncoordinated single-layer deception versus coordinated multilayer deception.
\end{itemize}

\section{Related work}

In this section, we review the key contributions in the areas of Cyber-Physical Systems (CPS), game modeling, interdependency analysis, deception techniques, and vulnerability-driven defenses. 
 
Guo et al.~\cite{guo2018stochastic} present a game-theoretic framework for modeling attacker-defender interactions in CPS as a continuous zero-sum Markov game. The attacker targets cyber-layer nodes to maximize damage, while the defender seeks to minimize it. An algorithm is proposed and validated through a case study of a smart grid. However, their model does not address attack propagation, multi-node targeting, cyber-layer interconnections, or physical-layer deception. Sanjab et al.~\cite{sanjab2016bounded} propose a CPS security model that uses game theory and cognitive hierarchy to account for differing reasoning abilities between attacker and defender. Their work assumes multi-node attacks but lacks attack propagation modeling, limiting applicability in complex, interconnected CPS environments. To analyze interdependencies in critical infrastructures, Akbarzadeh et al.~\cite{akbarzadeh2021identifying} introduce a Modified Dependency Structure Matrix (MDSM) that captures both interlayer and intralayer relationships. MDSM enables a compact representation and defines four parameters for assessing multi-order dependencies in large-scale systems.

Asghar et al.~\cite{asghar2024scalable} proposed scalable game-theoretic cyber deception strategies integrating both cyber and physical layers, formulating novel deception games and analyzing Nash equilibria. While their approach employs a double oracle framework with integer linear programming and a heuristic defender oracle achieving near-optimal defender payoffs within 16\% in large-scale simulations, which lacks empirical validation and scalability beyond simulation. Similarly, Yao et al.~\cite{yao2024bayesian} introduced a Bayesian-stochastic hybrid game model to defend industrial CPSs against persistent, uncertain attacks, effectively modeling cross-layer propagation but omitting deception mechanisms like honeypots or coordinated defense, leaving it vulnerable to adaptive attackers. In contrast, Sengupta et al.~\cite{sengupta2017game} examined web application defense through Moving Target Defense (MTD), modeled as a repeated Bayesian game that optimizes configuration switching based on CVE-derived metrics and vulnerability prioritization. However, their framework remains confined to the cyber layer, lacking the cross-layer coordination central to our proposed model.

\section{Problem Statement}

Defenders employ Unmanned Ground Vehicles (UGVs) to safeguard assets in the physical domain, with operations dependent on interconnected cyber-control systems, making both layers mutually vulnerable. Adversaries exploit this interdependence by targeting cyber nodes to disable physical defenses or by launching direct physical attacks \cite{asghar2024scalable}. In this multi-domain environment, our framework models a coordinated deception strategy where defenders disrupt adversarial reconnaissance across cyber and physical layers, while attackers conduct cross-domain intelligence gathering for a combined assault. This integrated scenario exemplifies how systematic, scalable defense strategies can be designed to address evolving cyber-physical threats in real-world contexts.

\begin{figure}
    \centering
    \includegraphics[width=8cm, height=4cm]{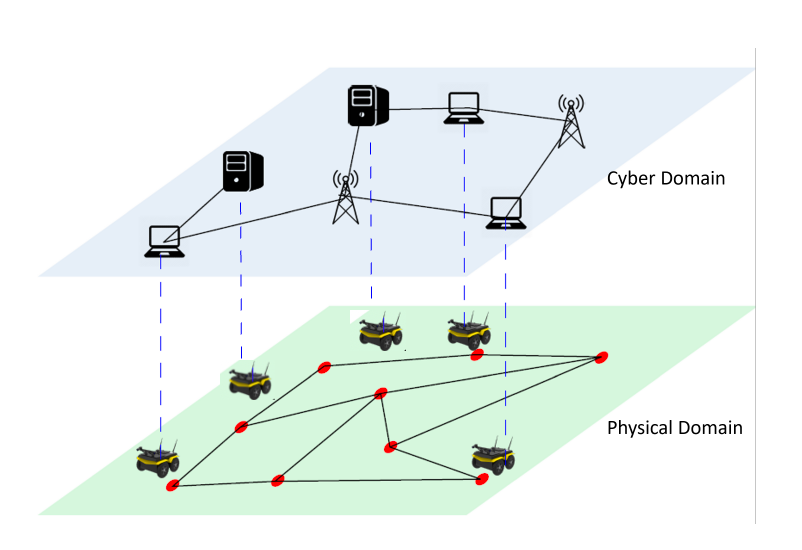}
    \caption{Multi-layer setting:  The blue dashed lines represent the replication of nodes across the cyber and physical layers.}
    \label{fig:ml_model}
\end{figure}

In multi-domain environments, deception is reinforced by mirroring vulnerabilities across cyber and physical layers. Limiting deception to a single layer risks exposing inconsistencies, enabling attackers to identify genuine assets. To address this, we create cyber replicas of physical components and physical replicas of cyber assets such as fake UGVs or sensors paired with decoy servers to obscure the true system topology and confuse adversarial reconnaissance (Fig.~\ref{fig:ml_model}). This coordinated cross-domain replication not only enhances deception effectiveness but also captures the intertwined complexity of cyber-physical dependencies. Since deception incurs asymmetric costs, with physical decoys being more expensive, the strategy’s success is measured by how effectively attackers engage with deceptive nodes and by achieving a higher payoff from coordinated deception than from isolated single-layer tactics. To optimize this sequential defender–attacker interaction, we formulate a bi-level Stackelberg game \cite{conitzer2006computing}, where the defender proactively deploys deception and the attacker subsequently conducts reconnaissance, allowing us to compute equilibrium strategies that maximize defensive utility.

\begin{table}[hbt!]
% \caption{The different known vulnerabilities in the defender's network, and the corresponding CVE ID with domain mentioned.}
\caption{List of vulnerabilities with CVE identifiers and domain classification.}
%\scriptsize
%\small
\tiny
\centering
\begin{tabular}{|l|l|l|l|l}
\hline
\makecell{\textbf{ID}} & \textbf{Vulnerability} & \textbf{CVE ID} & \textbf{Domain}\\
\hline
$a_{1}$ & Ripple20 & CVE-2020-10220 &  Cyber \\ \hline
$a_{2}$ & Heartbleed & CVE-2014-0160 &  Cyber \\ \hline
$a_{3}$ &  \makecell[l]{Microsoft SharePoint \\Remote Code Execution} & CVE-2020-16952 & Cyber \\ \hline
$a_{4}$ & \makecell[l]{Microsoft Exchange\\ Server Vulnerability} & CVE-2020-0688 & Cyber \\ \hline
$a_{5}$ & Log4Shell & CVE-2021-44228 & Cyber \\ \hline
$a_{6}$ & \makecell[l]{VeraEdge and Veralite\\ Unauthenticated \\Lua Code Execution} & CVE-2017-9389 & Physical \\ \hline
$a_{7}$ & \makecell[l]{Shekar Endoscope\\ Memory Corruption} & CVE-2017-10724 & Physical \\ \hline
$a_{8}$ & \makecell[l]{Siemens S7-1200/1500\\ PLC Vulnerability} & CVE-2019-10915 & Physical \\ \hline

\end{tabular}
\label{tab:cve_table}
\end{table}

We consider a system comprising five cyber nodes and three physical nodes, with each cyber node having a corresponding replica in the physical layer and each physical node mirrored in the cyber layer. Table~\ref{tab:cve_table} presents the list of nodes along with their associated CVE identifiers.

The CVSS score\footnote{https://nvd.nist.gov/} ranges from 0 to 10 and is divided into three categories: base, temporal, and environmental. The base score includes two subscores: (1) \textit{exploitability}, based on access vector ($AV$), access complexity ($AC$), and authentication (AU); and (2) \textit{impact}, reflecting potential harm to confidentiality ($C$), integrity ($I$), and availability ($A$). Full scoring details are available in the CVSS guide. Since CVSS v4 scores are not available for all CVEs, we utilize both CVSS v2 and v3 to demonstrate the generalization of our framework across different scoring versions.

In this work, we estimate the likelihood of exploitation using only the exploitability subscore, omitting impact and environmental metrics. The probability of successful exploitation for a given vulnerability \( e_i \) is computed following~\cite{poolsappasit2011dynamic} as:

\begin{equation} \label{eqn:cvss2}
Pr(e_i) = 2 \times AV \times AC \times AU
\end{equation}

Equation \ref{eqn:cvss2} represents the probability of a successful exploitation for a given vulnerability based on the corresponding CVSS metric version v2.

Additionally, CVSS version 3 introduces a scope metric, allowing differentiation when the exploit impacts components beyond the vulnerable system (e.g., privilege escalation). CVSS metric  v3 introduces the conditional
probability of vulnerability exploitation defined\cite{zhang2017conditional} in the following equation:

\begin{equation} \label{eqn:cvss3}
Pr(e_i) = AV \times AC \times PR \times UI
\end{equation}

Where $AV$ is access vector, $AC$ is access complexity, $PR$ is privilege required, and $UI$ is user interface.

% motivation of stakelberg game
\section{Game Model}

We represent the problem outlined earlier as a two-player, non-zero-sum normal form game, denoted as $\Gamma(\mathcal{N}, \mathcal{A}, \mathcal{R})$. Here, $\mathcal{N} = \{d, a\}$ represents the players, namely the defender and the attacker. The action sets of the players are represented by $\mathcal{A} = \{\mathcal{A}_d, \mathcal{A}_a\}$, and their corresponding reward functions are denoted as $\mathcal{R} = \{R_d, R_a\}$. The attacker's objective is to exploit vulnerabilities to compromise the system, while the defender's strategy involves deploying deception by mimicking vulnerabilities to mislead the attacker.

\subsection{Defender  Model}

Equation~\ref{eqn:vd} captures the loss of the defender when a specific vulnerability is exploited. The defender incurs a negative cost based on the probability of the exploit and impact of each vulnerability. In contrast, a failed exploit yields a positive reward, which increases for the physical layers compared to cyber layers.

\begin{equation} \label{eqn:vd}
V_{d}(i) = -1 * Pr(i) * IS_i + (1 - Pr(i)) * r_i  \quad \forall \, i \in A_{d}
\end{equation}

Where $Pr(i)$ is the probability of successful exploitation $i$, $IS$ is the impact score, and $r$ is the reward.

\subsection{Attacker Model}

Equation~\ref{eqn:va} represents the attacker’s gain from successfully exploiting specific vulnerabilities. If exploitation is successful, the attacker receives a reward based on the CVSS base score. In contrast, failed attempts result in a cost incurred by the attacker.

\begin{equation} \label{eqn:va}
V_{a}(j) = Pr(j) * BS_j - (1 - Pr(j)) * ES_j  \quad \forall \, j \in A_{a}
\end{equation}

Where $Pr$ is exploit probability, $BS$ is base score, and $ES$ is exploit score.

% To rescale a range between an arbitrary set of values [a, b] formula \footnote{\url{https://en.wikipedia.org/wiki/Feature_scaling}} given as,

\begin{align} \label{eq:minmax_norm}
x' = a + \left( \frac{x - x_{\min}}{x_{\max} - x_{\min}} \right) \cdot (b - a)
\end{align}

%where $x$ denotes the original value, $x_{min}$ and  $x_{max}$ represent the minimum and maximum values of the input list, $a$ and $b$ specify the new normalization bounds, and $x'$ denotes the resulting normalized value. 
The values of \( V_{d} \) and \( V_{a} \) were normalized to the intervals \([-10, -1]\) and \([1, 10]\), respectively, using Equation~\ref{eq:minmax_norm}.

Table ~\ref{tab:cve_table_exp} shows the exploit probability of various CVEs, along with the corresponding loss and gain values associated with defending or compromising the network. These metrics are analyzed across different CVSS versions, providing a comparative assessment of the risks and defensive priorities for each vulnerability.

\begin{table}[hbt!]
\caption{The known vulnerabilities in the defender's network, and the corresponding exploit probability in different cvss version.}
\label{tab:finetunned_model}
%\scriptsize
\centering
%\small
\tiny
\begin{tabular}{|c|c|c|c|c|c|c|}
\hline
%\toprule
\multicolumn{4}{|c|}{\textbf{CVSS V2}} & \multicolumn{3}{c|}{\textbf{CVSS V3}} \\  \hline
%\midrule
\textbf{ID} & \makecell[l]{\textbf{Exploit} \\ \textbf{probability}} & $\boldsymbol{V_{d}}$ & $\boldsymbol{V_{a}}$ & \makecell[l]{\textbf{Exploit} \\ \textbf{probability}} & $\boldsymbol{V_{d}}$ & $\boldsymbol{V_{a}}$\\ \hline
$a_{1}$ & 0.999 & -8.69 & 10 & 0.727 & -9.85 & 9.69\\ \hline
$a_{2}$ & 0.999 & -5.6 &  7.2 & 0.727 & -6.48 & 6.13\\ \hline
$a_{3}$ & 0.86 & -7.27 & 6.79 & 0.47 & -3.06 & 1.0\\ \hline
$a_{4}$ & 0.795 & -9.16 & 7.79 & 0.531 & -5.52 & 3.92\\ \hline
$a_{5}$ & 0.859 & -10.0 & 9.19 & 0.727 & -10.0 & 10\\ \hline
$a_{6}$ & 0.795 & -8.61 & 7.79 & 0.53 & -2.66 & 3.9\\ \hline
$a_{7}$ & 0.795 & -6.09 & 5.56 & 0.53 & -2.69 & 3.92\\ \hline
$a_{8}$ & 0.395 & -1.0 & 1.0 & 0.471 & -1.0 & 2.56\\ \hline

\end{tabular}
\label{tab:cve_table_exp}
\end{table}

\subsection{Rewards}

Recent work \cite{asghar2024scalable} consider zero-sum game formulation to calculate the value(significance) of nodes in computer network from one perspective such as defender or attacker. Alternatively, we consider non-zerosum game formulation where the defender and the attacker have different levels of preference for protecting the network from different vulnerabilities and exploits.

The defender reward is expressed as:

\begin{align} \label{eq:my_equation3}
      R_d(a_d, a_a) = Cap \cdot V_{a}(a_d)\cdot \mathbf{1}_{\left\{ a_d = a_a\right\}} & +Esc \cdot V_{d}(a_a)\cdot \mathbf{1}_{\left\{ a_d \neq a_a\right\}} \nonumber  \\ - C_d (a_d) + C_a (a_a)
\end{align}

The attacker reward is expressed as:

\begin{align} \label{eq:my_equation4}
      R_a(a_d, a_a) = Esc \cdot V_{a}(a_a)\cdot \mathbf{1}_{\left\{ a_d \neq a_a\right\}} & - Cap \cdot V_{a}(a_a)\cdot \mathbf{1}_{\left\{ a_d = a_a\right\}} \nonumber  \\ + C_d (a_d) - C_a (a_a)
\end{align}

\(Cap\) denotes the defender's reward upon capturing the attacker through interaction with a honeypot or fake UGV/sensor, whereas \(Esc\) represents the attacker’s gain from successfully exploiting vulnerabilities or evading deception. The deception cost \(C_d\) is proportional to the cyber and physical replicas deployed, with higher costs associated with physical layer replicas, while the attack cost \(C_a\) increases when targeting physical vulnerabilities. In Internet of Battlefield Things (IoBT) and military contexts, defending physical-layer vulnerabilities deserves higher priority, as their compromise can cause irreversible mission-critical failures, lacks continuous monitoring, and provides minimal forensic recovery capabilities~\cite{asghar2024scalable}.

\subsection{Stackelberg Game Model}

The reward function $R$ captures the strategic interaction between cyber and physical domains in our game-theoretic deception model. The defender (leader) deploys coordinated deception such as decoy systems and false signals to mislead the attacker, while the attacker (follower) observes and responds based on perceived system states. This sequential structure models proactive defense, where the defender’s strategy shapes the attacker’s observations. The Stackelberg equilibrium of the game yields the optimal deception policy that maximizes defender utility by invalidating attacker reconnaissance and reducing the likelihood of successful intrusions.

As we consider two player non-zero-sum game where $R_d$ $!=$ $R_a$. The defender’s target is to maximize the expected reward, whereas the attacker is the minimizing
agent. The expected utility of the defender can be expressed as: 

\begin{equation}
    U^d =A_d^T R_d A_a
\end{equation}

where $R_d$ is the game reward matrix of size $|A_d| \times |A_a|$.

The expected utility of the attacker can be expressed as: 
\begin{equation}
    U^a = A_d^TR_aA_a
\end{equation}

where $R_a$ is the game reward matrix of size $|A_d| \times |A_a|$.

We compute the Stackelberg equilibrium~\cite{conitzer2006computing} to maximize the expected utility of the defender, assuming that the attacker adopts an optimal response. The interaction is formulated as a MIQP, where the attacker’s pure strategy $A_a$ is modeled as a binary variable and the defender’s pure strategy $A_d$ represents honeypot deployment. The optimization determines the mixed strategy of the defender that ensures the best response of the attacker while satisfying all constraints of the game.

\begin{equation}
\label{eq:eqn_miqp}
\max_{i \in A_{a}} \quad U^d(i) \, A_{a_i}
\end{equation}

\begin{align}
\text{s.t.} \quad & A_{d_i} \in [0,1], \quad \forall \, i \in A_{d} \\
& A_{a_i} \in \{0,1\}, \quad \forall \, i \in A_{a} \\
&U^a(i) \,  A_{a_i} \geq U^a(i') \, A_{a_i} \quad \forall \, i, i' \in A_{a} \label{eq:bs}\\
& \sum_{i=1}^{|A_d|} A_{d_i} = 1 \label{eq:vp} \\
& \sum_{i=1}^{|A_a|} A_{a_i} = 1
\end{align}

In this formulation, the unknown variables include the defender's strategy $A_d$
and the attacker's action $A_a$. Equation ~\ref{eq:eqn_miqp} represents the objective function of the quadratic program (QP), aiming to maximize the defender’s expected utility. The inequality in Equation ~\ref{eq:bs} guarantees that the attacker adopts a best-response strategy, while Equation ~\ref{eq:vp} enforces that the defender’s strategy adheres to a valid probability distribution.

\subsection{Identifying the Most Strategically Critical Vulnerability}

In real-world cybersecurity, defenders face limited resources and cannot patch all vulnerabilities simultaneously~\cite{sengupta2017game}. Within our multi-domain deception framework, prioritizing which vulnerabilities to fix becomes complex due to the non-zero-sum nature of the game, where defender and attacker utilities are not strictly opposing. The defender must balance security gains and deception effectiveness across layers, identifying vulnerabilities whose patching or deception yields the greatest strategic benefit considering cross-layer dependencies and attacker incentives.

\begin{algorithm}
\caption{Finding the Most Critical Vulnerability}
%\small
\scriptsize
\label{alg:critical_vuln}
\textbf{Input:} Utility \\
\textbf{Output:} $a^*$ \\
\textbf{Result:} Finds the most critical vulnerability that, when fixed, yields the highest defender utility.
\begin{algorithmic}[1]
\State $max\_defender\_utility \gets -\infty$
\ForAll{$a \in A_{a}$}
    \State $A_{a}' \gets A_{a} \setminus \{a\}$
    \State $obj\_val \gets$ Solve MIQP (10) with action set $A_{a}'$
    \If{$obj\_val > max\_defender\_utility$}
        \State $max\_defender\_utility \gets obj\_val$
        \State $a^* \gets a$
    \EndIf
\EndFor
\State \Return $a^*$
\end{algorithmic}
\end{algorithm}

Building on Equation~\ref{eq:eqn_miqp}, we identify the attack action \(a \in \mathcal{A}_a\) whose removal yields the maximum defender utility gain. This is done by iteratively removing each attack, reformulating the MIQP, and recomputing defender utility. The attack whose exclusion provides the greatest increase is deemed the most critical vulnerability, as summarized in Algorithm~\ref{alg:critical_vuln}.

\section{Result}

\begin{table}[h!]
%\centering

%\bottomrule

\tiny
%\scriptsize
%\caption{Comparison of Strategies under CVSS v2.0 and CVSS v3.0}
\begin{tabular}{|c|c|c|c|c|}
%\toprule

\hline
\multirow{2}{*}{\textbf{Method}} & \multicolumn{2}{c|}{\textbf{CVSS v2}} & \multicolumn{2}{c|}{\textbf{CVSS v3}} \\
\cline{2-5}
 & \textbf{Mixed Strategy} & \makecell{\textbf{Defender's} \\ \textbf{Utility}} & \textbf{Mixed Strategy} & \makecell{\textbf{Defender's} \\ \textbf{Utility}} \\
\hline
URS & \makecell{(.125, .125, .125, .125,\\ .125, .125, .125, .125)} & -60.25 & \makecell{(.125, .125, .125, .125,\\ .125, .125, .125, .125)} & -87.0 \\
\hline
GS & \makecell{(0, 0, 0, 0,\\ 0, 0, 0, 1)} & -58.0 & \makecell{(0, 0, 0, 0,\\ 0, 0, 0, 1)} & -102.0 \\
\hline
SG & \makecell{(0.27, 0.12, 0.09, 0.16,\\ 0.24, 0.11, 0.0, 0.0)} & \textbf{-52.1} & \makecell{(0.385, 0.22, 0.0, 0.0,\\ 0.394, 0.0, 0.0, 0.0)} & \textbf{-48.94} \\
\hline
\end{tabular}
\caption{Comparison between the strategies generated by URS, GS, and SG for CVSS v2 and v3}
\label{tab:strategy_comparison}
\end{table}

We evaluate the proposed Stackelberg game-based framework (SG) against two baselines Uniform Random Strategy (URS) and Greedy Strategy (GS) to assess its effectiveness in strategy generation and multilayer deception. The cyber and physical layers are defined with corresponding vulnerabilities and CVSS-based exploit probabilities to compute defender and attacker rewards. Cyber-to-physical replicas incur higher deployment costs, making cost-aware optimization crucial. Using MIQP, we compute the Stackelberg equilibrium~\cite{conitzer2006computing} to refine defender and attacker strategies and analyze how SG outperforms URS and GS in coordinated deception and critical vulnerability identification across the cyber-physical domain.

\begin{figure}[h!]%
    \centering
    \begin{subfigure}{.50\columnwidth}
        \includegraphics[width=\columnwidth, height=4cm]{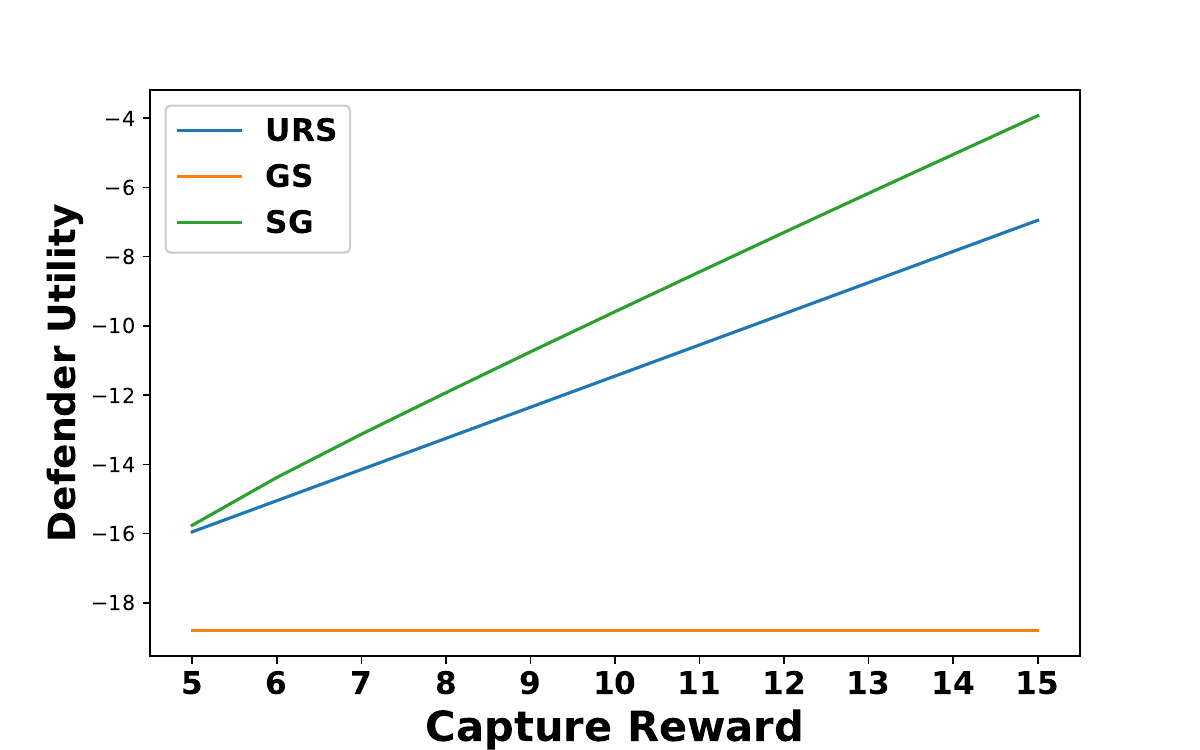}
    \end{subfigure}\hfill%
    \begin{subfigure}{.50\columnwidth}
        \includegraphics[width=\columnwidth, height=4cm]{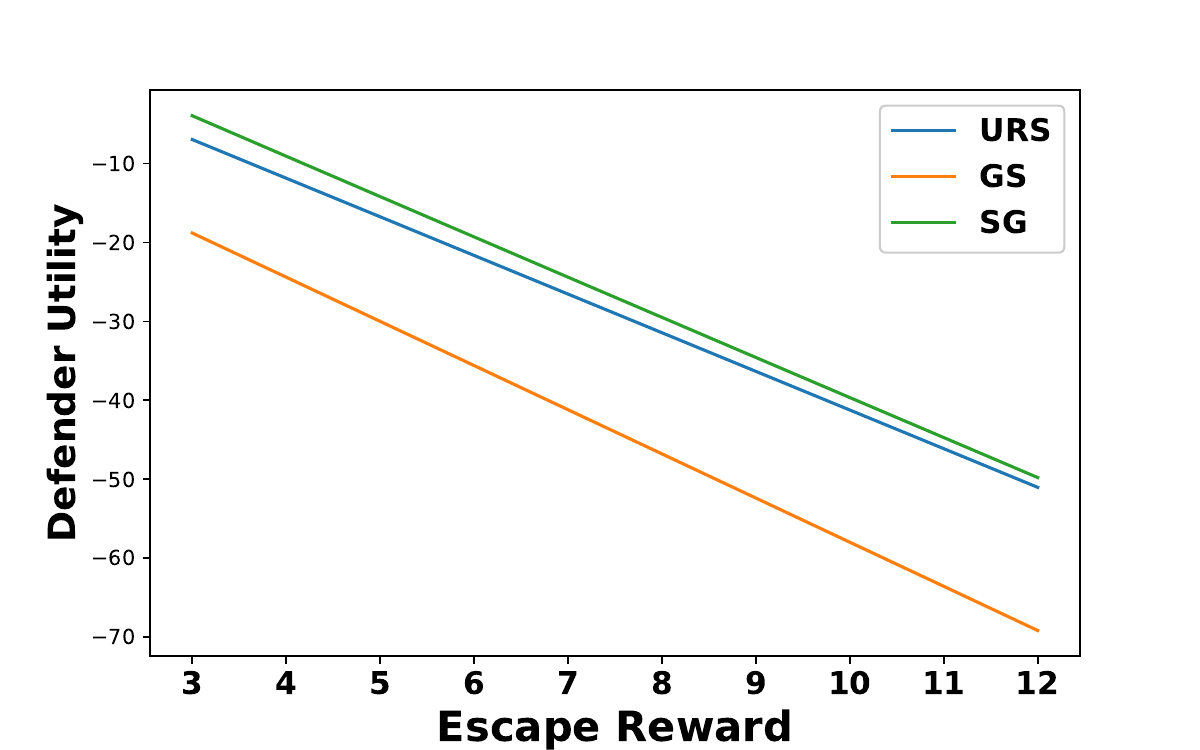}
    \end{subfigure}\hfill%
    \caption{Defender reward over different Cap and Esc values for CVSS v2.}%
    \label{fig:capesc2}%
\end{figure}

\begin{figure}[h!]%
    \centering
    \begin{subfigure}{.50\columnwidth}
        \includegraphics[width=\columnwidth, height=4cm]{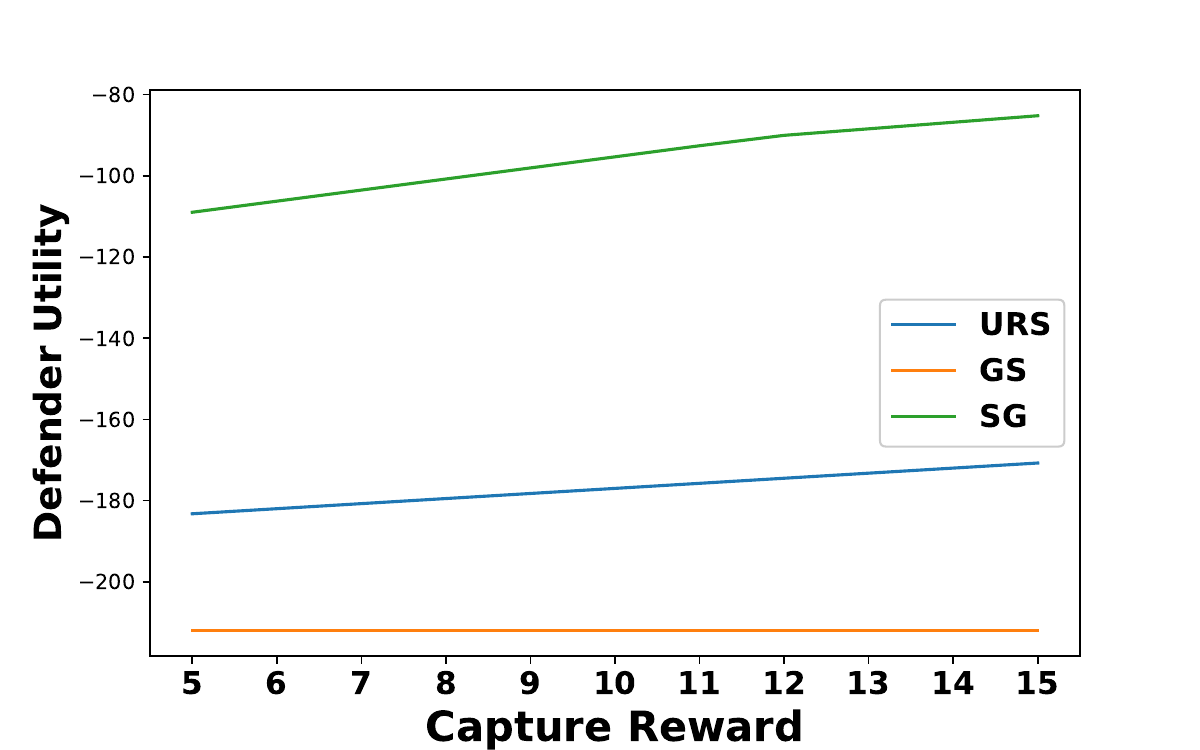}
    \end{subfigure}\hfill%
    \begin{subfigure}{.50\columnwidth}
        \includegraphics[width=\columnwidth, height=4cm]{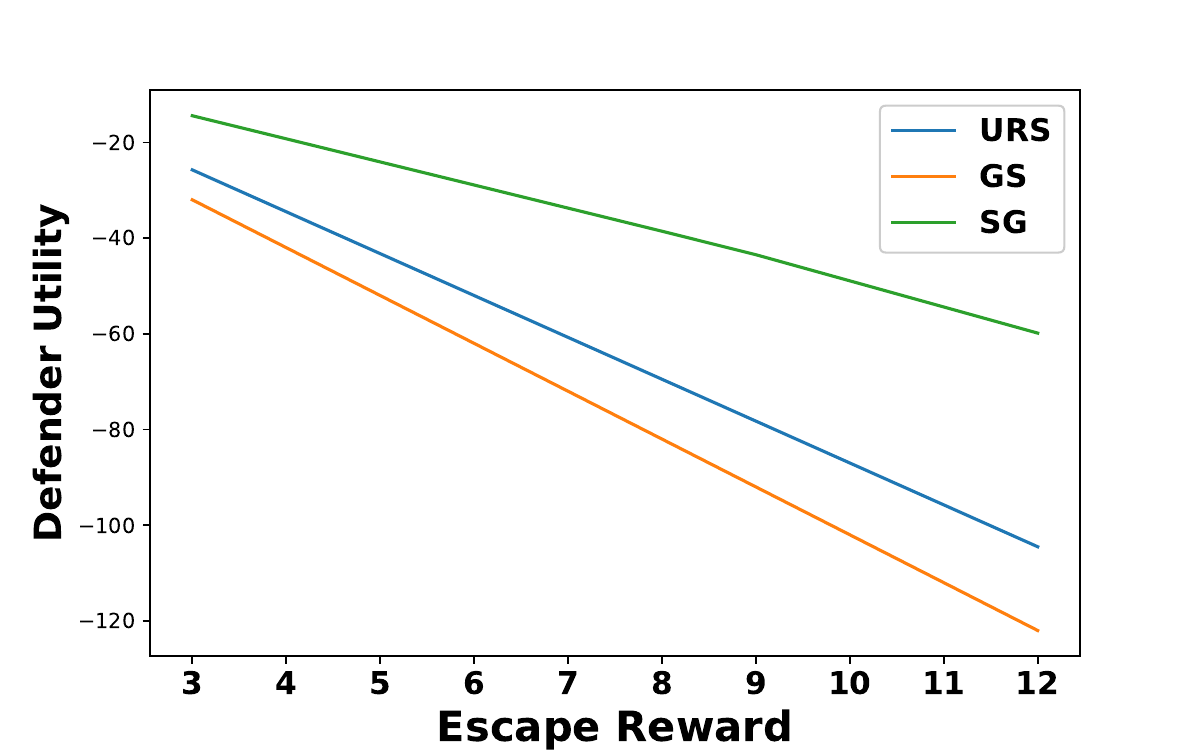}
    \end{subfigure}\hfill%
    \caption{Defender reward over different Cap and Esc values for CVSS v3.}%
    \label{fig:capesc3}%
\end{figure}

Table~\ref{tab:strategy_comparison} shows that the SG model achieves the highest defender utility across both CVSS v2 and v3 scoring schemes, outperforming URS and GS strategies. URS maintains deception uncertainty by distributing defenses uniformly when attacker behavior is unknown, whereas GS efficiently prioritizes high-impact nodes to minimize immediate losses. In multi-domain settings, the defender favors coordinated deception on cyber-layer nodes due to cost and strategic leverage. Figures~\ref{fig:capesc2} and~\ref{fig:capesc3} further confirm that SG consistently yields superior defender utility across varying capture and escape rewards, with utility increasing linearly with capture gain and decreasing with attacker escape advantage.

\begin{figure}[h!]
    \centering
    \includegraphics[width=8.6cm, height=4cm]{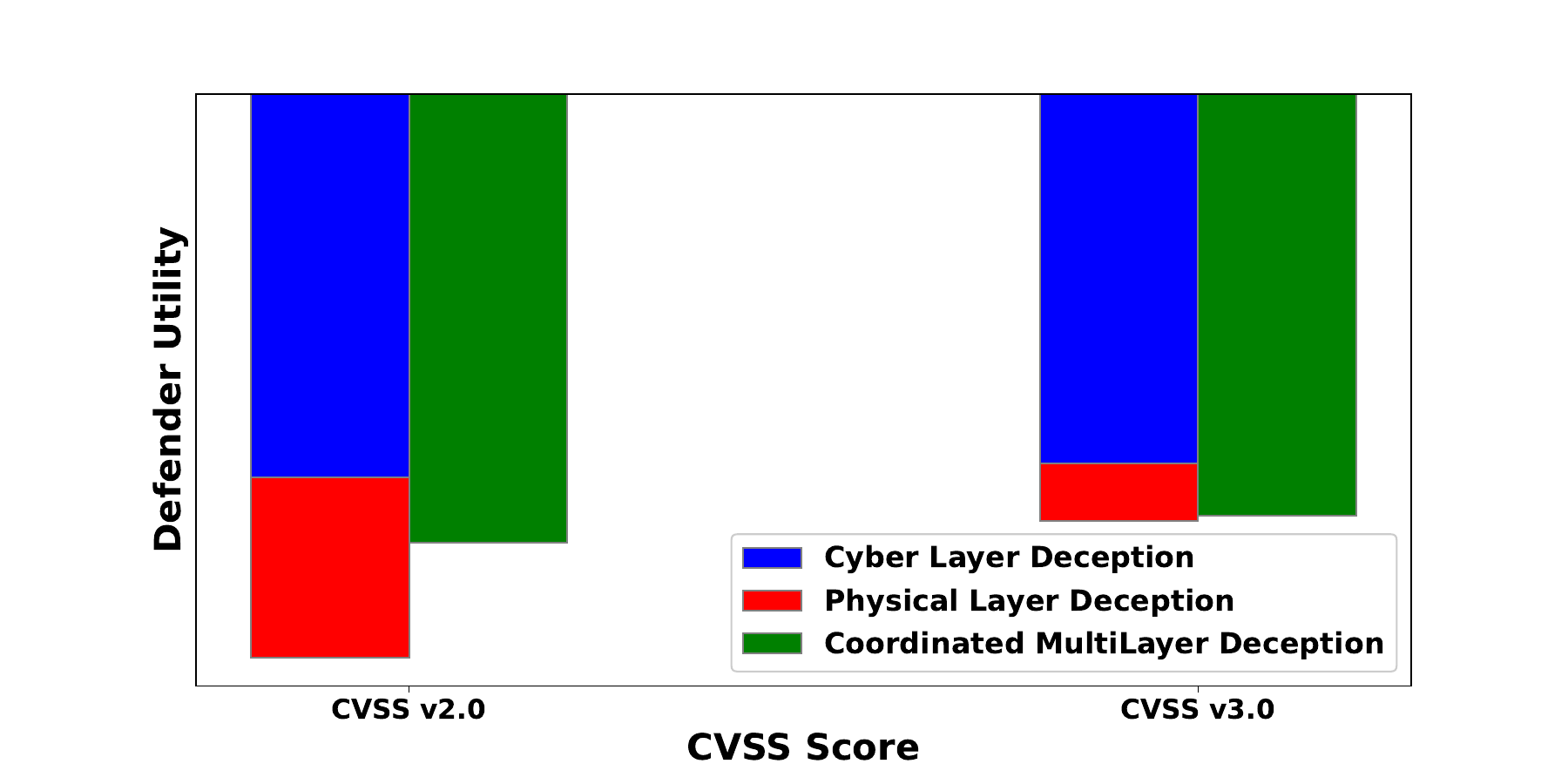}
    \caption{Single vs Coordinated Deception}
    \label{fig:def_util_compare}
\end{figure}

Figure~\ref{fig:def_util_compare} shows that coordinated multilayer deception achieves the highest defender utility across both CVSS v2 and v3 scoring, outperforming cyber-only and physical-only strategies. This demonstrates the benefit of integrating deception across domains. The utilities obtained by removing one vulnerability at a time are: for CVSS v2, 
$\langle a_1 : -46.4;\ a_2 : -53.8;\ a_3 : -50.1;\ a_4 : -48.6;\ a_5 : -47.04;\ a_6 : -50.48;\ a_7 : -52.1;\ a_8 : -52.1 \rangle$, 
and for CVSS v3, 
$\langle a_1 : -26.59;\ a_2 : -28.49;\ a_3 : -48.94;\ a_4 : -48.94;\ a_5 : -26.49;\ a_6 : -48.94;\ a_7 : -48.94;\ a_8 : -48.94 \rangle$. 
Under CVSS v2, vulnerability \(a_1\) yields the greatest improvement in defender utility when removed, while under CVSS v3, \(a_5\) is most impactful, confirming the model’s ability to identify high-priority vulnerabilities for targeted mitigation.

\section{Conclusion}

This paper explores the role of coordinated deception, emphasizing its integration across cyber and physical layers to enhance defensive strategies. By leveraging a Stackelberg game model, we analyze the interaction between attackers and defenders, optimizing deception strategies using CVSS-based reward functions derived from the NVD. Our framework demonstrates how multi-layered deception can effectively disrupt attacker reconnaissance and enhance threat detection. Comparative analysis with state-of-the-art strategies highlights the practical applicability and security benefits of our approach, offering a scalable and systematic method for strengthening cyber-physical system resilience.

\bibliographystyle{IEEEtran}
\bibliography{references}

\end{document}